\documentstyle[preprint,aps,psfig]{revtex}

\begin{document}

\draft

\title{Jamming transition in a cellular automaton model for traffic flow}

 \author{B.\ Eisenbl\"atter$^a$, L. Santen$^b$, A.\ Schadschneider$^b$, 
M.\ Schreckenberg$^a$\\}
\address{$^a$ Fachbereich 10, Gerhard-Mercator-Universit\"at Duisburg,
  47048 Duisburg, Germany}

\address{$^b$ Institut f\"ur Theoretische Physik, Universit\"at zu K\"oln,
50937 K\"oln, Germany}

\date{\today}

\maketitle

\begin{abstract}
\noindent
The cellular automaton model for traffic flow exhibits a jamming
transition from a free-flow phase to a congested phase. In the
deterministic case this transition corresponds to a critical point
with diverging correlation length. In the presence of noise, however,
no consistent picture has emerged up to now. We present data from
numerical simulations which suggest the absence of critical behavior.
The transition of the deterministic case is smeared out and one
only observes the remnants of the critical point.
\end{abstract}

\vskip 0.5 cm

\section{Introduction}
The cellular automaton (CA) approach to traffic flow  theory
\cite{NaSch} has attracted much interest in recent years (see e.g.\
\cite{juelbook}). Compared to the earlier attempts in modeling
traffic flow (see e.g.\ \cite{juelbook,Light,Kern,helbing} and references
therein) CA models can be used very 
efficiently for computer simulations. This makes it possible to perform 
real time simulations even for very large networks \cite{network,city}.
Due to the relevance of these models for applications it is important to
understand the underlying physics thoroughly.

The cellular automaton model of Nagel and Schreckenberg (NaSch model) 
\cite{NaSch} provides a simple but quite realistic description of traffic flow.
The road is divided into $L$ cells so that the model is discrete in space 
and time. Each cell can either be empty or occupied by one of
$N$ cars $j=1,\ldots, N$ with velocities $v_j=0,...,v_{max}$. $v_{max}$ 
is assumed to be the same for all the cars.
The update is divided into four steps which are applied {\em in parallel} to
all cars. The first step (R1) is an acceleration step. The velocities $v_j$
of each car $j$ not already propagating with the maximum velocity $v_{max}$ 
are increased by one. The second step (R2) is designed to avoid accidents.
If a car has $d_j$ empty cells in front of it and its velocity (after 
step (R1)) exceeds $d_j$ the velocity is reduced to $d_j$. Up to now 
the dynamics is completely deterministic. Noise is 
introduced via the randomisation step (R3). Here the velocities of moving cars 
$(v_j\geq 1)$ are decreased by one with probability $p$. 
The steps (R1)--(R3) give the new velocity $v_j$ for each car $j$. 
In the last step of the update procedure the positions $x_j$ of the 
cars are shifted by $v_j$ cells (R4) to $x_j+v_j$. 
We consider here only periodic boundary conditions. Thus the model
contains three parameters: the maximum velocity $v_{max}$, the 
probability for braking $p$, and the average density $\rho =N/L$. 

\begin{figure}[h]
\centerline{\psfig{figure=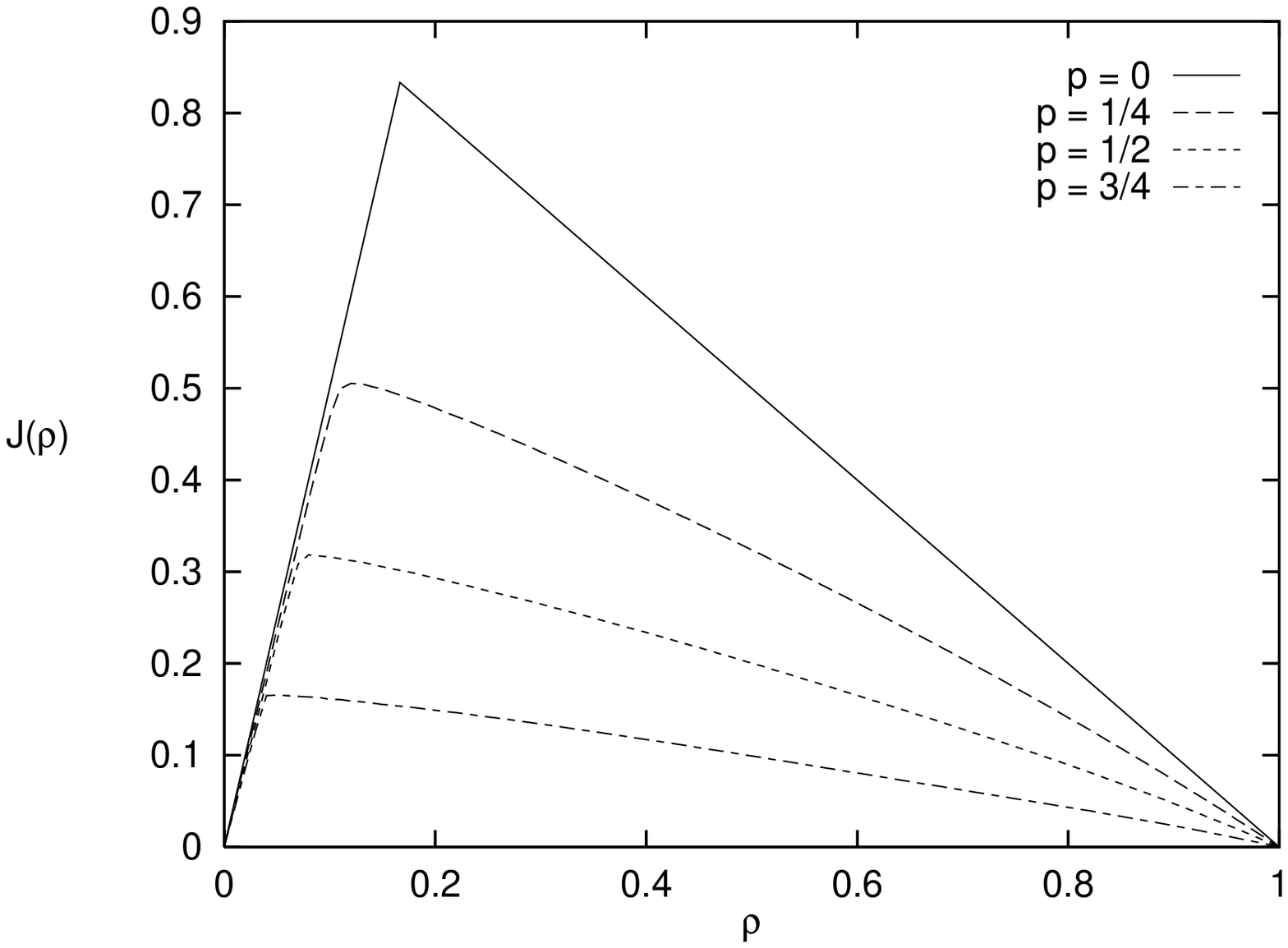,bbllx=0pt,bblly=0pt,bburx=600pt,bbury=450pt,height=8cm}}
\caption{\protect{Fundamental diagram for different values of $p$ 
($v_{max} =5$).}\hfill}
\label{fig1}
\end{figure}

A basic feature of traffic models is the relation between
density $\rho$ and the average flow $J=\rho \bar v$ (fundamental diagram)
where $\bar v=\frac{1}{N}\sum_{j=1}^Nv_j$ is the average velocity. 
Fig.\ \ref{fig1} shows the fundamental diagram
for different values of $p$. One observes two effects as the noise $p$
is increased, namely a decrease of the flow and a shift of the maximum 
towards smaller densities. In the low-density limit $\rho \ll 1$ one 
always finds free flow behavior with $J(\rho) \simeq (v_{max}-p)\rho$ 
whereas for high densities $1-\rho \ll 1$ one has $J(\rho) 
\simeq (1-p)(1-\rho)$.

Although the model cannot be solved exactly for arbitrary parameter values,   
two limits of the model can be treated analytically. Firstly the
case $v_{max} = 1$ which is solved excactly with improved mean field
methods \cite{Cluster,Comf}. Here the fundamental diagram is 
symmetric due to particle-hole symmetry. Considering larger maximum 
velocities $v_{max}\geq 2$ one can obtain solutions only in the
deterministic limit $p=0$ where the flow is given by
\begin{equation}
  \label{flow_det}
  J(\rho) = \min ( \rho~v_{max} , 1-\rho ).
\end{equation}
In the free flow regime where $J(\rho)=\rho v_{max}$ all cars propagate with 
maximum velocity whereas in the jammed phase the flow is limited by the 
number of empty cells. These limits will be discussed further in the 
later sections.

Such a transition from a free flow regime at low densities to a congested 
flow regime where start and stop waves dominate the dynamics are quite
typically for traffic flow.  Several attempts have been made to 
explain the nature of this transition in the CA model 
\cite{Vilar}-\cite{Luebeck}. 
It seems, however, that no consensus has been reached yet.
Here we present results of an extensive numerical investigation of the
parameter  dependence of the transition in the NaSch model. 
We examine several quantities which give information about the location
$\rho_c$ and the nature of the phase transition.

The outline of the paper is as follows: In the next section we discuss
the relaxation into the steady state. Section 3 is devoted to
measurements of an order-parameter. Section 4 shows the behavior of 
the spatial correlation function. Our results are
discussed in the final section.

\section{Relaxation}

A characteristic feature of a second order phase transition is the
divergence of the relaxation time at the transition point. For 
technical reasons Cs\'anyi and Kert\'{e}sz \cite{Csanyi}
made no direct 
measurements of the relaxation time, but used the following approach:
Starting from a random configuration of cars with velocity $v_j=0$ the
average velocity $\bar v(t)$ is measured at each time step $t$. 
For $t\rightarrow\infty$ the system reaches a stationary state 
with average velocity $\left < \bar v_{\infty} \right >$. The
relaxation time is characterised by the parameter \cite{Csanyi}
\begin{equation}
  \label{eq:relax}
  \tau= \int\limits_{0}^{\infty}\biggl[ {\rm min}\{v^{\ast}(t),\ 
  \langle \bar{v}_{\infty}\rangle\} - \langle \bar{v}(t)\rangle\biggr]\  
  dt \ . 
\end{equation}
$v^{\ast}(t)$ denotes the  average velocity  in the acceleration phase
$t\rightarrow 0$ for low vehicle density $\rho\rightarrow
0$. Because the vehicles do not  interact with each other, 
$v^{\ast}(t) = (1-p)t$ holds in this regime.
Thus the relaxation time is obtained by summing up the deviations of the
average velocity $\langle \bar{v}(t)\rangle $ 
from the values of a system with one single vehicle which can move
without interactions with other cars ($\rho\rightarrow 0$).
One finds a maximum of the relaxation parameter near, but {\em below}, the
density of maximum flow for  $p=0.25$ and ${v_{\mathit{max}}}=5$ 
(see \cite{Csanyi}). 

Within this investigation we extended the set of braking parameters 
($p = 0 \dots 0.75$) in order to study the parameter dependence of the
maximum of $\tau$. We took into account  system sizes up to $L=30000$
where the position $\rho_c$  of the maximum of $\tau$ becomes size
independent. The transition density is given by $\rho_c$ of the
largest system we took into account.

\begin{figure}[h]
  \centerline{\psfig{figure=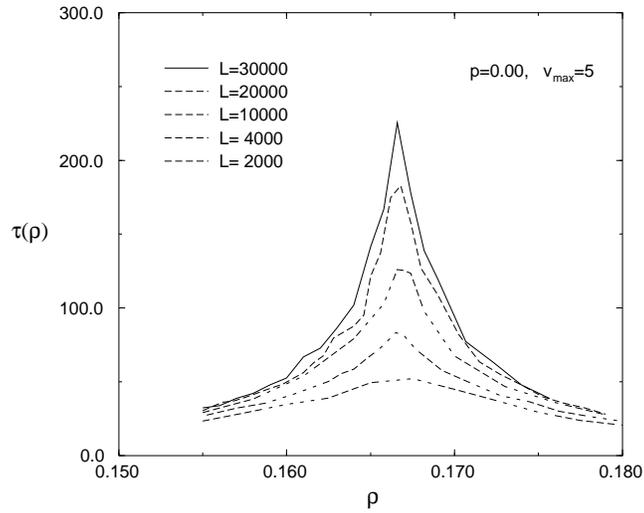,bbllx=40pt,bblly=70pt,bburx=600pt,bbury=550pt,height=8cm}}
  \caption{\protect{Relaxation parameter near the transition density
    for different system sizes ($v_{max} =5, p =0$). }}
\label{fig2}
\end{figure}

\begin{figure}[h]
\centerline{\psfig{figure=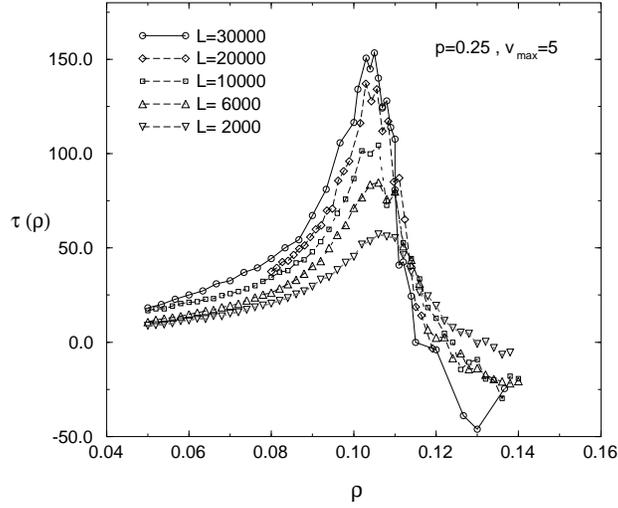,bbllx=0pt,bblly=0pt,bburx=600pt,bbury=490pt,height=8cm}}
  \caption{\protect{Relaxation parameter near the transition density
    for a higher value of the braking probability ($v_{max} =5, p =0.25$). }}
\label{fig3}
\end{figure}

\begin{figure}[h]
 \centerline{\psfig{figure=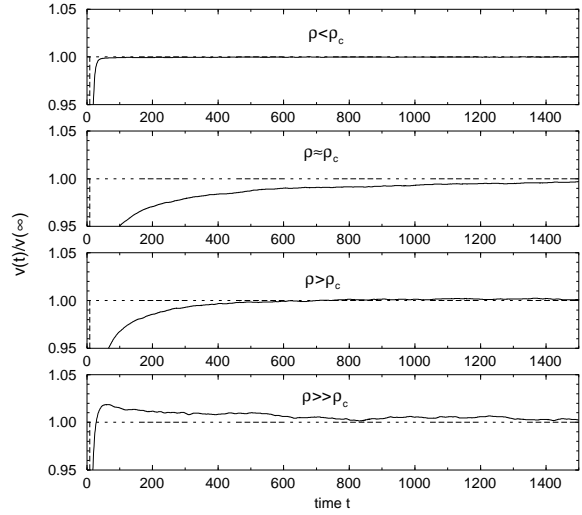,bbllx=0pt,bblly=200pt,bburx=600pt,bbury=800pt,height=8cm}}
 \caption{\protect{Time dependence of the average velocity. After a
      few time steps the average velocity reaches its absolut maximum.}}
\label{fig4}
\end{figure}

The results for $p=0$ and $p=0.25$ are shown in Figs.\ \ref{fig2}
and \ref{fig3}. A comparison of $\rho_c$ with the density of maximum flow
$\rho({q_{\mathit{max}}})$ shows smaller values of the transition
densities for all values  of $p$ taken into account. 
Taking the magnitude of $\tau$ as a characteristic value for the
relaxation time one can estimate the dynamical exponent. Furthermore 
the scaling behavior of the width  $\sigma(L)$ and height $\tau_m(L)$
of the peak has been taken into account,
\begin{equation}
  \label{eq:defz}
  \tau_m(L)\propto L^z,
\qquad
  \label{eq:defnu}
  \sigma(L)\propto L^{-1/\nu}.
\end{equation}
We find $z=0.28$ and $\nu =5.7$ for $p=0.5$ and $z=0.36$ and $\nu
=6.8$ for $p=0.25$.
Note that the peak is not symmetric so that it is difficult to 
determine its width.
Comparing  our data with \cite{Csanyi} two facts have to be mentioned.
First, our results for $p=0.25$ are completely different from the exponents
obtained in \cite{Csanyi}. The second  remarkable point is the
occurance of negative values for the relaxation times (see Fig.\
\ref{fig3}). 
This effect is not shown in \cite{Csanyi}. One can think that it emerges from
inaccurate measurements or finite size effects, but  
the negative values  are  a consequence of the definition  (\ref{eq:relax}).
If we look at the time evolution of  
$ \left < \bar{v}(t)\right >/ \left <\bar v_{\infty} \right >$
we see  the reason for this unpleasent feature (Fig.\ \ref{fig4}): 
For $\rho>\rho_{c}$ the system gets temporarily into states which have
a higher average velocity than the stationary state such that    
$\left < \bar{v}(t)\right > > \left < \bar v_{\infty} \right>$ holds
within this time interval.
This over-reaction is a consequence of the relaxation mechanism which can
be divided into two phases for $p>0$. Within the first few time steps
small clusters  which occur in the initial configuration vanish.
The second phase is characterized by the growth of surviving
jams. More and more cars get trapped into large jams and therefore the
average flow decreases to its stationary value. This decrease causes 
negative values of $\tau$ at large densities.

Finally one should note that (\ref{eq:relax}) can only be interpreted
as a relaxation time for a purely exponential decay, 
$\langle \bar{v}_{\infty}\rangle - \langle \bar{v}(t)\rangle 
\propto e^{-t/\tau}$. Figure \ref{fig4} shows, however, that this is
not the case for $\rho>\rho_c$, where one even finds a non-monotonic
relaxational behavior. In order to get a  clear picture of the nature
of the transition one should therefore examine various quantities.

\section{Order Parameter}
For a proper description of the transition one should introduce an
order parameter which has a qualitative different behavior within
the two phases. A first candidate would be the analogue of the magnetisation 
in the Ising model, i.e.\ the number of cars. However, since this quantity is
conserved in the NaSch model it can not serve as an order parameter. 
Therefore the density of nearest neighbour pairs
\begin{equation}
  \label{opdef}
   m = \frac{1}{L}\sum_{i=1}^{L} n_in_{i+1},
\end{equation}
with $n_i=0$ for an empty cell and $n_i=1$ for a cell occupied by a
car (irrespective of its velocity), is the simplest choice of a local 
quantity with a nontrivial behavior at the transition density. Taking into 
account the braking rule (R2) $m$ gives the density of those cars with 
velocity $0$ which had to brake due to the next car ahead.
Although the order parameter introduced in \cite{Vilar} 
is defined as a time average it shows a quite similar behavior. For large 
time periods it measures the densities of cars with velocity 0 \cite{eisi}.
Vilar et al.\ \cite{Vilar} only investigated the deterministic
case $p=0$ for which their order parameter is identical to ours, but
also in the presence of noise the values differ only slightly. 
First we will discuss the behavior of the order parameter in the case 
$p=0$. Below the transition density,
\begin{equation}
  \label{rhocdet}
  \rho_c = \frac {1}{v_{max}+1},
\end{equation}
the order parameter vanishes because every car has at least $v_{max}$ 
empty sites in front and propagates with $v_{max}$. Within the jammed 
phase the flow is
limited by the number of empty cells and also stopped cars occur. 
In the presence of noise the behavior of the order
parameter qualitativly changes in the vicinity of the transition
density. Within this region $m$ decays exponentially. Assuming $m$ is
a possible choice for the order parameter this implies the absence of
criticality in the nondeterministic case.
Figure \ref{fig6} shows that the order parameter does not exhibit
a sharp transition. Although it becomes rather small for small
densities it is still different from zero. The situation is quite similar
to the behavior of the order parameter in finite systems \cite{finite}.
The transition is smeared out by the noise and the transition density
is shifted towards smaller values.
In order to have a suitable criterion for the determination of the transition 
density, we analysed the scaling behavior of the order parameter near the 
transition density $\rho_c$. Fig.~\ref{fig7} shows that one gets a quite 
reasonable data collaps using the scaling form
\begin{equation}
  \label{rhoscale}
   \overline m(\rho) = \Pi(p) m(\rho + \Delta \rho_c).
\end{equation}
$\Pi(p)$ is a scaling factor and $\Delta \rho_c$  is the shift of
the transition density compared to the deterministic value (\ref{rhocdet}).
The values of the transition densities are shown in Fig.\ \ref{fig8}. 
This results are in good agreement with the results
obtained from the measurement of $\tau$.

\section{Spatial Correlations}
A striking feature of second order phase transitions is the
occurence of a diverging length scale at criticality and a corresponding
algebraic decay of the correlation function.
 Using lattice gas variables the density-density correlation function 
is defined by
\begin{equation}
  \label{korrdef}
  G(r) = \frac{1}{L}\sum_{i=1}^{L} n_in_{i+r}- \rho^2 .
\end{equation}
Again it is very instructive to consider the deterministic case
($p=0$) first. In the vicinity of the transition density one oberves a
decay of the amplitude of $G(r)$ for larger values of the distance
between the sites. Precisely at $\rho_c$, however, the correlation 
function is given by
\begin{equation}
  \label{korrdet}
   G(r) = \cases{\rho_c-\rho_c^2 & for\ $r \equiv 0$ mod$(v_{max}+1)$  \cr
-\rho_c^2  & else \cr}
\end{equation}
because there are exactly $v_{max}$ empty sites in front of each car.
Considering small, but finite, values of $p$ the correlation function
has the same structure as in the deterministic case, but the amplitude
decays exponentially for all values of $\rho$.
The decay of the amplitude determines the correlation length for a
given pair of $(p, \rho)$, which is finite for all densities in the
presence of noise.
 The maximal value of the correlation length $\xi_{max}$
determines the transition density for small values of $p$. 
Numerically we find 
\begin{equation}
  \label{ximaxasym}
  \xi_{max} \sim p^{-\frac{1}{2}}.
\end{equation}
In fact, this picture can be confirmed analytically for $v_{max}=1$ 
\cite{unp}. Using the results of \cite{Cluster} one obtains $\xi_{max}^{-1}
=\ln\left(\frac{1-p}{1-p-\sqrt{p}}\right)$ for the correlation length 
$\xi_{max}$ at $\rho=1/2$. Therefore $\xi_{max}\propto 1/\sqrt{p}$ for 
small $p$.
This exponent seems to be  independent of $v_{max}$ although the 
particle-hole symmetry is broken for  $v_{max}>1$.
If one considers larger values of $p$ the correlation length gives not
the relevant length scale which is then determined by the size distribution 
of jams. A numerical analysis of this limit is
quite difficult and has to be referred to future work. 
\section{Discussion and Summary}
Our results suggest a consistent picture of the jamming transition in the
NaSch CA. Measurements of the order parameter and the
correlation function show that critical behavior only occurs in the
deterministic limit where the transition density is given by 
$\rho_c = (v_{max}+1)^{-1}$ (see also \cite{Vilar,NaHerr}). The presence of 
any noise destroys long-range correlations. The behavior is 
analogous to a second order phase transition in finite systems \cite{finite}. 
We have, however, checked carefully that our results are not affected by 
finite-size effects and are solely due to the presence of noise.

Analogous behavior is also found in the Ising chain in a transverse field
\cite{ferenc}. The transverse field $\Gamma$ is the control parameter
and corresponds to the density $\rho$ in the NaSch model whereas the 
temperature $T$ corresponds to the noise parameter $p$. This correspondance
can be used to predict scaling laws. These predictions are currently under
investigation and results will be published elsewhere.

We found qualitatively the same behavior of the relaxation parameter 
as shown in \cite{Csanyi}, but some  
important new features have been observed. An important result is the
occurance of negative values of $\tau$ which is a consequence of the
relaxation mechanism beyond the transition density:  Within
the first few time steps small jams which are present in the initial
condition die out. The second phase is dominated by the formation of
large jams. Thus at a certain time intervall the average flow is
systematically larger than the stationary value, which causes negative
contributions at that time. Consequently one has to question
whether $\tau$ gives meaningful results concerning the relaxation time or
not. 

The order parameter does not vanish exactly, but the transition density
could be determined from the scaling behavior. This suggests that the
system is not critical in a strict sense. 
Measurements of the density-density correlation function and the 
correlation length confirm this picture. We find a finite
correlation length in the presence of noise ($p>0$). The maximum 
correlation length diverges in the deterministic 
limit like $\frac{1}{\sqrt p} $ for all values of $v_{max}$ we investigated
($v_{max}=2,3,5$). For the case $v_{max}=1$ this result
can also be confirmed by analytical calculations.

Our conclusions have to be compared with those of other investigations where
signals of a second order transition also in nondeterministic cases
have been found. From our point of view these results are 
either a consequence of a special limit considered or the 
methods chosen. Nagel and Paczuski \cite{Pac} showed the existence
of self-organized critical behavior for the outflow region of a 
large jam in the cruise-control limit. They found a scale-invariant size 
distribution of jams from measurements far downstream of the megajam.
In this region most of the cars propagate without any fluctuations
such that this limit is also an example for scale-invariance in
deterministic flow. Very recently an investigation of the probabilistic
version of the NaSch model has been performed \cite{Luebeck} and it has 
been argued, that at the jamming transition critical behavior occurs also 
for the nondeterministic cases. 
However, the order parameter introduced in \cite{Luebeck} does also not
vanish exactly below the transition density. All the data presented in
\cite{Luebeck} are consistent with our interpretation of the nature
of the transition. In contrast to the view of \cite{Luebeck} we expect
true phase separation only in the limit $p\to 1$.

Another indication for the absence of critical behavior is the 
well-established fact (see e.g.\ \cite{Csanyi} and Fig.\ \ref{fig8})
that the density of maximum flow $\rho(q_{max})$ and the transition 
density $\rho_c$ are different.
It would be rather surprising if the system exhibits a genuine
second order phase transition with diverging correlation length.
Correlations obviously favor states with higher flow (see e.g.\ Figs.\
\ref{fig9} and \ref{fig10}, which show that occupancies of cells in front
of a car are surpressed which is the generalization of the particle-hole
attraction observed in \cite{Cluster} for $v_{max}=1$). Therefore
one should expect that the state with the strongest correlations is also
the state with the highest flow, as in the deterministic case.

In conclusion, we found the absence of criticality in the NaSch model
in the presence of noise. For finite $p$ the second order transition 
of the deterministic case is smeared out, similar to the situation of 
a second order transition in a finite system. Here this effect
is caused by the presence of noise, $p>0$.
For small values of $p$ one finds an
ordering transition close to $\rho_c=1/(v_{max}+1)$. Larger values of
the noise favour the formation of jams and a tendency to phase
separation occurs (see also \cite{ZPR,Luebeck,Chowd}). We therefore
are currently investigating the limit $p \rightarrow 1$ more carefully.
The results will be presented in a future publication.\\[0.3cm]

\noindent{\bf Acknowledgments:}\\
Part of this work has been performed within the research program of
the SFB 341 (K\"oln--Aachen--J\"ulich). We like to thank D.\ Chowdhury,
F.\ Igl\'oi, and J.\ Kert\'esz for valuable discussions and the
the HLRZ at the Forschungszentrum J\"ulich for generous allocation
of computing time on the Intel Paragon XP/S~10.

\begin{figure}[h]
  \centerline{\psfig{figure=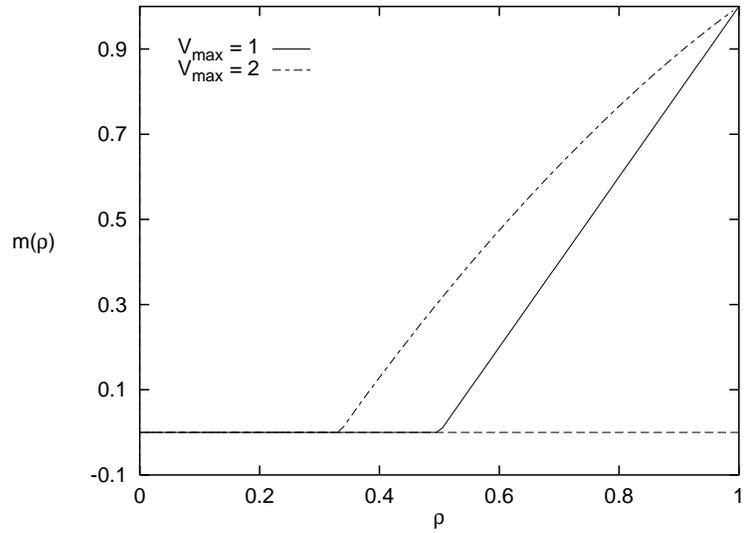,bbllx=0pt,bblly=0pt,bburx=600pt,bbury=400pt,height=8cm}}
  \caption{\protect{Order parameter for the deterministic model
      ($v_{max}= 1,2$). Below the transition density $m$
      vanishes exactly.}}
\label{fig5}
\end{figure}

\begin{figure}[h]
 \centerline{\psfig{figure=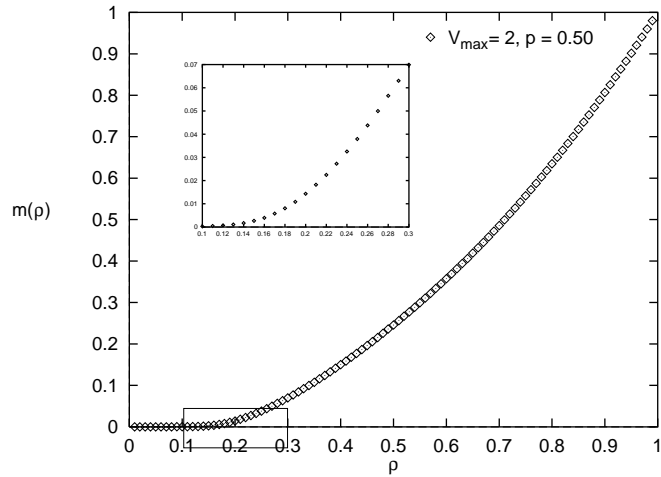,bbllx=50pt,bblly=125pt,bburx=600pt,bbury=550pt,height=8cm}}
 \caption{\protect{Behavior of the order parameter for a finite
      braking probability. It does not vanish exactly for $\rho < \rho_c$
      but converges smoothly to zero even for small values of the braking 
      probability $p$.}}
\label{fig6}
\end{figure}

\begin{figure}[h]
  \centerline{\psfig{figure=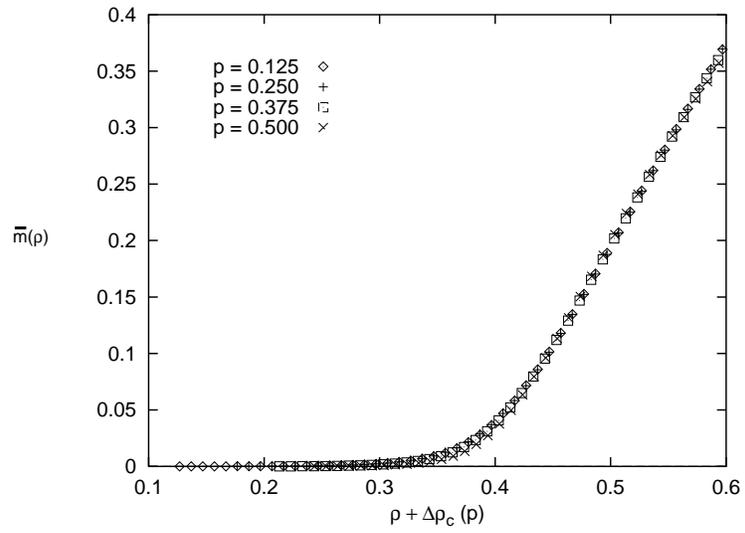,bbllx=50pt,bblly=110pt,bburx=600pt,bbury=500pt,height=8cm}}
  \caption{\protect{Scaling-plot of the order parameter. In the
      vicinity of the transition density one gets a reasonable data
      collaps. The density shift determines the transition density
      for a given $p$.}}
\label{fig7}
\end{figure}

\begin{figure}[h]
  \centerline{\psfig{figure=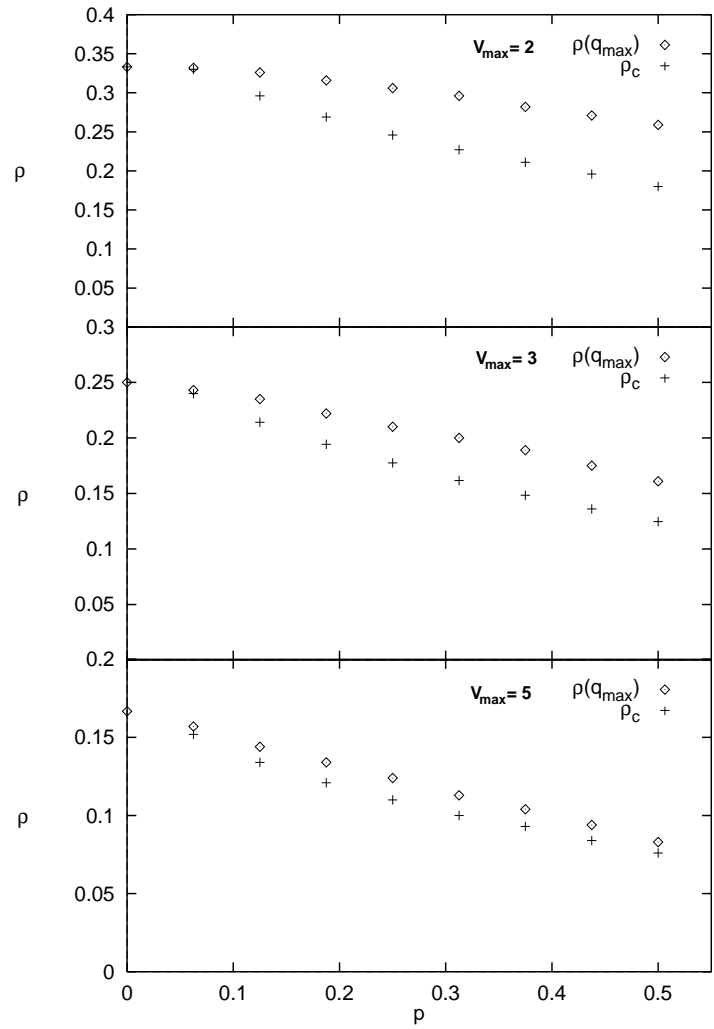,bbllx=0pt,bblly=0pt,bburx=600pt,bbury=760pt,height=15cm}}
  \caption{\protect{Comparison between transition density and density
      of maximum flow.}\hfill}
\label{fig8}
\end{figure}

\begin{figure}[h]
  \centerline{\psfig{figure=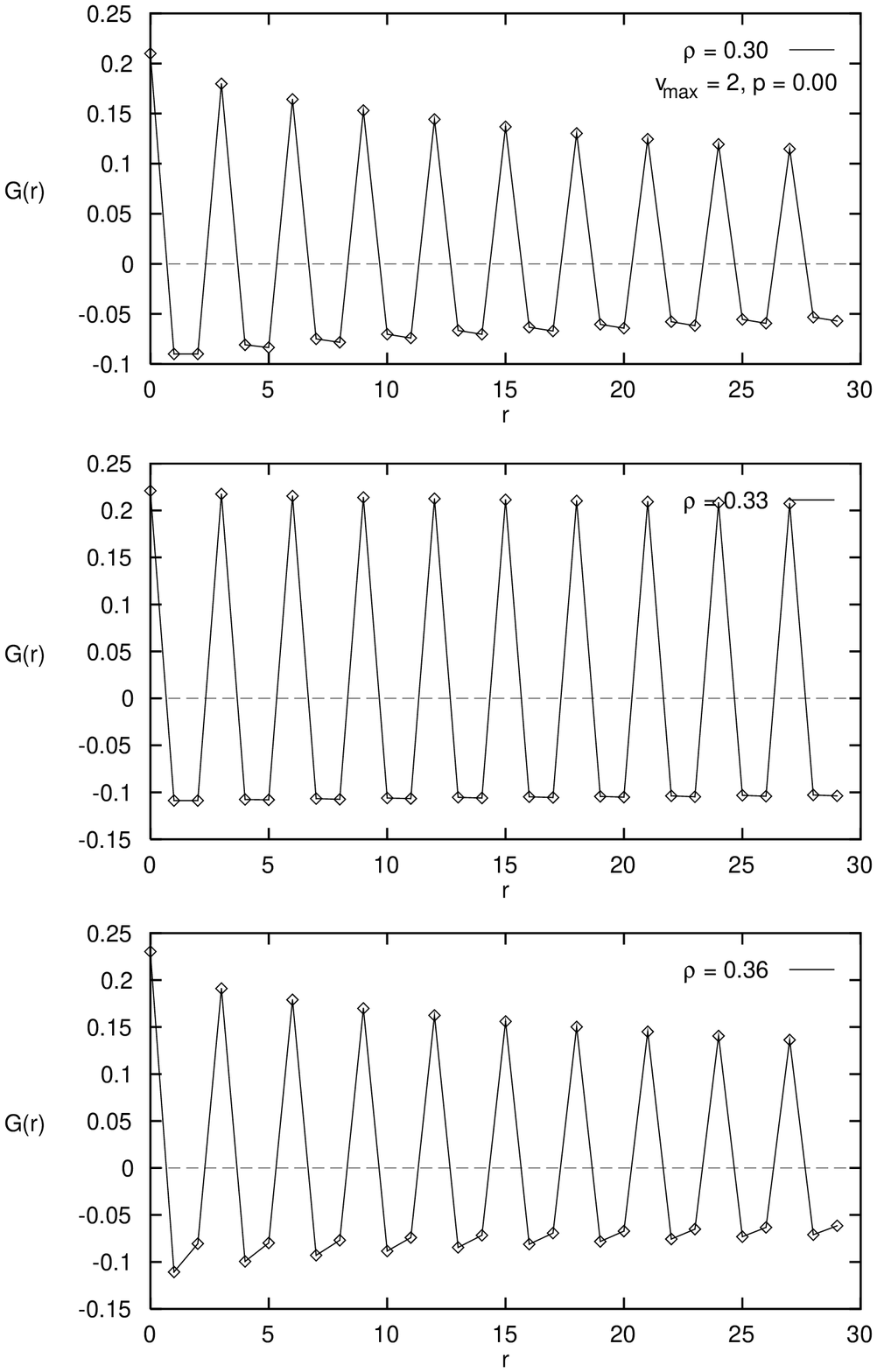,bbllx=50pt,bblly=30pt,bburx=550pt,bbury=730pt,height=12cm}}
  \caption{\protect{Correlation function in the vicinity of the phase
      transition for the deterministic limit. At $\rho = \rho_c$ the
      amplitude is independent of the distance $r$.}}
\label{fig9}
\end{figure}

\begin{figure}[h]
  \centerline{\psfig{figure=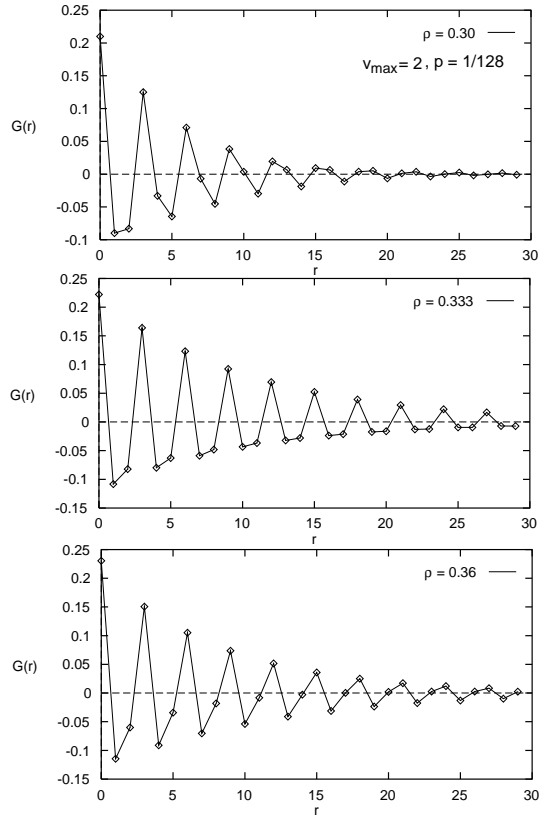,bbllx=70pt,bblly=20pt,bburx=550pt,bbury=720pt,height=12cm}}
  \caption{\protect{Correlation function in the presence of noise. The
      amplitude of the correlation function decays exponentially for
      all values of $\rho$.}}
\label{fig10}
\end{figure}

\begin{figure}[h]
  \centerline{\psfig{figure=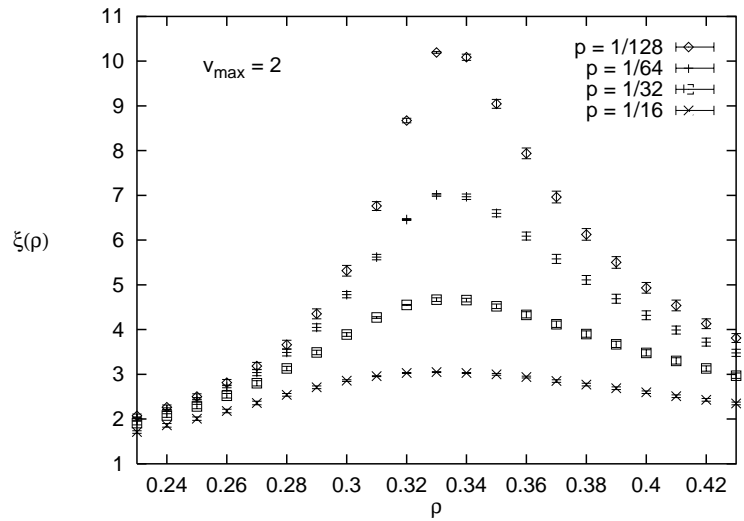,bbllx=20pt,bblly=0pt,bburx=600pt,bbury=400pt,height=8cm}}
  \caption{\protect{Density dependence of the correlation length in
      the vicinity of the transition density.}}
\label{fig11}
\end{figure}

\begin{figure}[h]
   \centerline{\psfig{figure=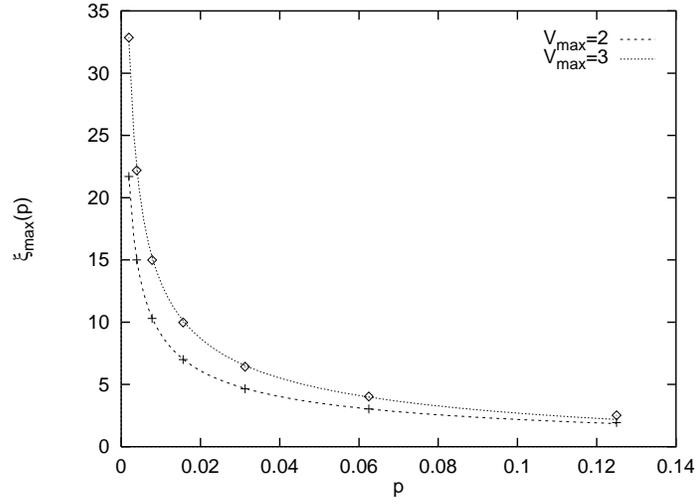,bbllx=20pt,bblly=20pt,bburx=560pt,bbury=450pt,height=8cm}}
 \caption{\protect{Noise dependence of $\xi_{max}$ for different
      maximum velocities. Independent of the maximum velocity 
      $\xi_{max}(p) \sim 1/ \sqrt{p}$ holds in the limit $p
      \rightarrow 0$. }}
\label{fig12}
\end{figure}


\begin{references}
\bibitem{ferenc} We like to thank F.\ Igl\'oi for pointing this out to us.
\end{references}
\end{document}